\documentclass[%
reprint,
amsmath,amssymb,
pra,
]{revtex4-2}
\usepackage{graphicx}
\usepackage{dcolumn}
\usepackage{bm}
\usepackage[english]{babel}
\usepackage{enumitem}

\usepackage{xcolor}
\usepackage{hyperref}
\hypersetup{
	colorlinks=true,
	linkcolor=blue,    
	citecolor=blue,   
	filecolor=magenta, 
	urlcolor=blue      
}

\begin{document}

\preprint{APS/123-QED}

\title{Electrically controlled focusing of de Broglie matter waves by Fresnel zone plate}

\author{Sushanta Barman}
\email{sushanta@iitk.ac.in}

\author{Sudeep Bhattacharjee}%
\email{sudeepb@iitk.ac.in}
\affiliation{Department of Physics, Indian Institute of Technology-Kanpur, Kanpur 208016, India}%

\begin{abstract}
The evolution from classical optics to quantum matter wave optics has been influenced by transformative optical devices like Fresnel zone plates (FZP). This study investigates the focusing of helium atoms by an electrically biased FZP, made of chromium (Cr), using numerical solutions of the time-dependent Schrödinger equation (TDSE) with the split-step Fourier method. The \(n\)th opaque zone of the FZP is subject to electrostatic biasing using three methods: (i) \(V_n = V_1\), where \(V_1\) is the biasing voltage applied to the central zone ($n=1$), (ii) \(V_n = V_1 \sqrt{n}\), and (iii) \(V_n = V_1 \sin (k_E n)\), with \(k_E\) being the radial modulation factor in the biasing voltage. The effects of electrostatic biasing on the transmitted wave packet distribution, transmission coefficient (\(T_c\)), focal length (\(f\)), size of the focused wave packet (\(\sigma_F\)), and focusing efficiency (\(\eta\)) are analyzed. Results show that electrical biasing modulates diffractive focusing via induced polarization potential. Biasing with \(V_n = V_1\) reduces the wave packet transmission and focusing efficiency. Biasing with \(V_n = V_1 \sqrt{n}\) enhances \(T_c\) by \(\sim 38\%\), \(f\) by \(\sim 73.5\%\), and \(\eta\) from 6\% to 20\%. Biasing with \(V_n = V_1 \sin(k_E n)\) offers better control, achieving a focusing efficiency of $\sim 10.5\%$. These findings suggest new methods for higher efficiency focusing of matter waves, with applications in helium microscopes, cold atom trapping, and high-precision atom lithography for quantum electronic devices.

\end{abstract}

\maketitle


\section{\label{introduction}Introduction}
The historical emergence of Fresnel zone plates (FZP) can be traced back to the early 19th century when the field of wave optics was pioneered by Augustin-Jean Fresnel \cite{soret1875ueber}. The zone plates are designed with alternating transparent and opaque concentric rings. The width of each ring is precisely adjusted to delay the wave passing through it by half a wavelength \cite{hecht2012optics}. This innovation revolutionized wave optics by achieving finer resolution and precise focusing beyond conventional lenses \cite{Tong_2023,PhysRevB.102.024420, mohacsi2017interlaced, PhysRevLett.96.127401, PhysRevB.74.033405, deng2017carbon}. Over the period of time, FZPs have enhanced various fields, including microscopy \cite{kubec2022achromatic,PhysRevLett.99.264801, sanli20183d, PhysRevLett.103.180801}, imaging \cite{rana2023three, PhysRevA.83.043808}, and spectroscopy \cite{marschall2017transmission, PhysRevE.105.065207, doi:10.1021/acs.nanolett.9b01970, qiao2022ultrasensitive, PhysRevA.97.053806}. They have also been employed in focusing acoustic waves \cite{tarrazo2019acoustic} and spin-waves \cite{doi:10.1021/acsnano.0c07076, PhysRevB.102.024420, sluka2019emission, albisetti2018nanoscale}, in the phase manipulation of electromagnetic waves using metasurfaces \cite{chen2018phase, s23084137}, in optical tweezers \cite{cheng2016fractal}, terahertz imaging \cite{nano13142037, app12157788}, astrophotonics \cite{Anand_2021}, and in nano and integrated photonics \cite{Flenner:20, Geints_2021, D2NR00594H}. Similarly, in atom optics, neutral atomic matter waves have been focused using nano-structured FZPs, applying principles similar to those used in classical optics \cite{PhysRevLett.67.3231, PhysRevLett.83.4229}. The FZP provides universal focusing, applicable even to fragile, reactive, and excited atomic species \cite{PhysRevLett.83.4229}. The focusing occurs without the physical interaction of particles with the zone plate structure as they pass through open annuli, a consequence of quantum mechanical principles and the wave-like properties of atoms.

Commercial particle microscopes using high-energy charged particles can cause sample charging and damage non-conductive samples, posing challenges for imaging fragile and insulating materials like polymeric nanostructures and biological specimens \cite{Eder_2012}. Alternatives such as atomic force microscopy \cite{RevModPhys.75.949} or scanning near-field optical microscopy \cite{Fleischer_2012_313_338} are not suitable for high aspect ratio structures \cite{doi:10.1021_acsnano.8b07216}. Low-energy neutral beams, particularly helium atoms, offer minimal surface damage due to their inertness and non-penetrating nature in the low energy range ($\leq100$ meV) \cite{PALAU2023113753}. Moreover, intense and neutral helium beams can be generated easily through supersonic expansion, matching the momentum and energy of single-crystal surfaces for efficient probing. Consequently, significant efforts have been made to focus low-energy He beams for the development of neutral atom microscopy \cite{PhysRevLett.67.3231, PhysRevLett.83.4229, Eder_2012, M_Koch_2008, REHBEIN2000685, 10.1063/1.5143950, 10.1116_1.2987955, Judd_2010, HATCHWELL2024113951, BERGIN2022113453, Bergin2020}, with a comprehensive overview presented in a recent review \cite{PALAU2023113753}.

In 1991, Carnal et al. achieved the initial focusing of neutral He atoms using a negative FZP (opaque central zone) \cite{PhysRevLett.67.3231}. Metastable helium atoms with various de Broglie wavelengths ($\lambda_{dB}=0.5 - 2.5$ \AA) were imaged through a 210 $\mu$m wide zone plate, demonstrating the potential for imaging two-dimensional sub-micron structures. This milestone paved the way for the development of an atomic microscope. Subsequently, helium atomic beams have been focused to a size of $\leq 2.0$ $\mu$m with a demagnification factor of 0.4 using the FZP \cite{PhysRevLett.83.4229, REHBEIN2000685}, and submicron focusing has been achieved through subsequent developments \cite{Eder_2012}. In 2008, the first microscopy image using neutral He atoms as the probe was captured with FZP focusing \cite{M_Koch_2008}. Moreover, FZPs have also been employed to focus deuterium molecules \cite{10.1116_1.2987955} and Bose-Einstein condensates of alkali atoms \cite{Judd_2010}, facilitating erasable atom lithography for quantum electronic devices. Recently, reflection images were taken using a FZP in a helium atom microscope \cite{FLATABO2024113961}.

Despite advancements, the resolution of an atom microscope is found to be limited by the outermost zone width \cite{PhysRevLett.83.4229, PhysRevA.91.043608}. Fabricating FZPs with nanometer precision is impractical due to low-energy atomic beams. Atom sieves with multiple pinholes, aligned with Fresnel zones, overcome this, reducing the fabrication limit to 3-5 nm \cite{PhysRevA.91.043608}. A zeroth-order filter or order-sorting aperture addresses signal contrast issues, allowing low-intensity measurements \cite{PhysRevA.95.023618}. Consequently, optimizing the microscope design is crucial to maximize intensity for a given resolution and outermost zone width \cite{PhysRevA.95.013611}. 

Given the importance of manipulating low-energy neutral atoms \cite{Judd_2010, BELL1998587, 10.1063/1.2172412}, as indicated by the above studies, investigating novel methods using FZPs to enhance focusing efficiency is crucial. Advances in the manipulation of atomic and matter waves have enabled precise control and imaging with unprecedented capabilities. Fujita et al. \cite{PhysRevLett.84.4027} demonstrated the use of electric fields in atomic holography to modulate reconstructed atomic images in real time, achieving dynamic control through precise electric field application. Shimizu's review \cite{SHIMIZU200073} further highlights the foundational principles of atom holography, emphasizing the critical role of coherent atomic beams and their interaction with holographic structures in producing complex atomic patterns. Recent experiments have demonstrated that electric field control of quantum reflection for matter waves using periodically microstructured surfaces can be achieved \cite{PhysRevA.95.043639}. Specifically, external biasing voltage modifies the interaction between the surface and matter waves, suggesting a method for electrically tunable reflection. However, no studies have yet explored the manipulation of matter waves by an electrostatically biased FZP or how such biasing influences focusing properties, which could lead to new applications in quantum technologies.

\begin{figure*}
	\centering
	\includegraphics[scale=0.85]{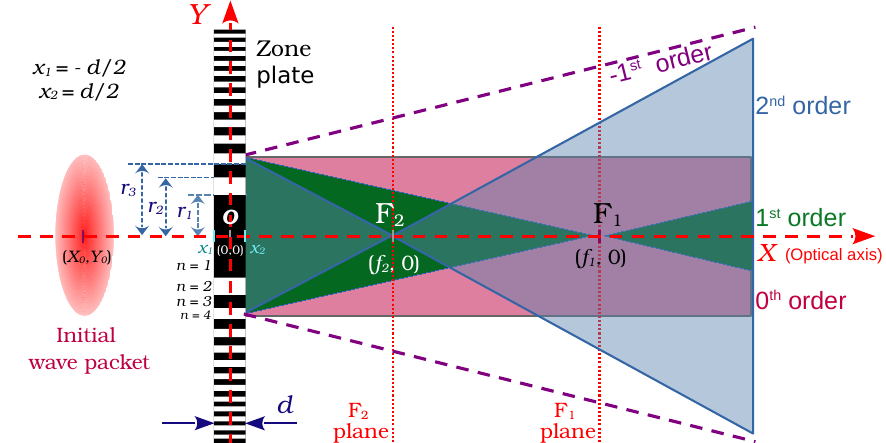}
	\caption{A schematic diagram of a Fresnel zone plate (FZP) made of chromium (Cr) with a width \(d\) (=50 nm) is employed to focus helium atoms. The odd zones (odd values of \(n\)) are opaque, and the even zones (even values of \(n\)) are transparent, defining it as a negative zone plate with the central zone (\(n=1\)) being opaque. The focal points, labeled as F\textsubscript{1} and F\textsubscript{2}, correspond to the first- and second-order focusing, respectively. After transmission through the FZP, the unfocused part of the transmitted wave packet is referred to as the zeroth-order, while the defocused part is characterized by the negative orders.}
	\label{schematic_diagram_zone_plate}
\end{figure*}

In this study, the focusing of cold He atoms using an electrically biased FZP constructed from chromium (Cr) is investigated numerically. The negative-type FZP, featuring an opaque central zone (\(n=1\)) and odd \(n\)-indexed zones, is subjected to DC biasing in three configurations: (i) \(V_n = V_1\), where \(V_1\) is the biasing voltage applied to the central zone, (ii) \(V_n = V_1 \sqrt{n}\), and (iii) \(V_n = V_1 \sin (k_E n)\), with \(k_E\) being the radial modulation factor in the biasing voltage. These configurations are inspired by an analogy with optical FZPs, where phase modulation across the zones is critical for focusing \cite{10.1119/1.18587}. The sinusoidal case is motivated by its relevance to optical systems, whereas the \(\sim \sqrt{n}\) case arises from considerations of phase modulation, analogous to the radial variation of zones in FZPs. Here, the applied biasing schemes act to control phase shifts induced on the matter waves, akin to the role of kinoform zone plates in achieving high efficiency in optics \cite{10.1119/1.18587}. The radially modulated and non-uniform voltage patterns provide precise control over wavefront modulation, enabling adaptable and efficient focusing behavior. The impact of biasing on the transmitted wave packet distribution, transmission coefficient (\(T_c\)), focal length (\(f\)), size of the foc used wave packet (\(\sigma_F\)), and focusing efficiency (\(\eta\)) is analyzed. It is observed that biasing with \(V_n=V_1\) decreases transmission and focusing efficiency, while biasing with \(V_n=V_1 \sqrt{n}\) enhances transmission by 38\%, increases focal length by 73.5\%, and improves efficiency from 6\% to 20.2\%. Biasing with \(V_n = V_1 \sin(k_E n)\) provides better control through the parameters \(k_E\) and \(V_1\), achieving a focusing efficiency of 10.5\%. This study provides insights into manipulating matter wave transmission through FZPs and their focal properties using external electrical biasing. Moreover, our numerical framework incorporates Casimir-Polder forces, demonstrating their impact on atom-surface interactions and the potential of electrical biasing to dynamically control aberrations and enhance the focusing performance of FZPs. This study also lays the groundwork for optimizing zone plate designs for matter-wave focusing, which could benefit from advancements in nanofabrication techniques to achieve nanometer-scale edge smoothness.

The article is structured as follows: Section \ref{methods} explains the theoretical methods. Section \ref{result_and_discussions} presents and discusses the simulation results. Finally, Section \ref{conclusions} draws the conclusions.
\section{\label{methods} Theoretical Methods}
The Hamiltonian ($\textbf{H}$) for a He atom interacting with an electrically biased metallic FZP can be expressed as:
\begin{align}
	\textbf{H} = -\frac{\hbar^2}{2m} \bm{\nabla}^2 + V_g(\textbf{r}) +  V_{int}(\textbf{r}) +  V_{induced}(\textbf{r}),
	\label{hamiltonian}
\end{align}
where $m$ and $\hbar$ are the mass of a helium atom and reduced Planck's constant, respectively. $V_g(\textbf{r})$ is the geometrical potential energy arising from the opaque zones of the FZP. $V_{\text{int}}(\mathbf{r})$ describes the potential energy arising from the long-range Casimir-van der Waals interaction between the He atom and the metallic surface of the FZP. Finally, $V_{\text{induced}}(\mathbf{r})$ represents the induced potential energy resulting from the polarization of He atoms in the presence of an electric field.

\subsection{The geometrical potential energy ($V_g(\textbf{r})$)}
In this study, a negative FZP is considered, as employed in previous studies \cite{PhysRevLett.67.3231, PhysRevLett.83.4229,Eder_2012, PhysRevA.95.023618} for better mechanical stability. The alternating opaque and transparent zones are depicted in Fig. \ref{schematic_diagram_zone_plate}. A negative FZP starts with an opaque zone at the center (taken as the first zone), and all other odd $n$th zones are made opaque. The radius of the $n$th zone, denoted as $r_n$, is the distance from the center of the FZP to its outer edge, as depicted in Fig. \ref{schematic_diagram_zone_plate} and given by $r_n=\sqrt{n \lambda_{dB} f}$ \cite{hecht2012optics, PhysRevLett.67.3231}, where $n \in \mathbb{Z}^{+}$, $\lambda_{dB}$ represents the de Broglie wavelength of the particle, and $f$ denotes the focal length of the FZP. The geometric function for a negative FZP can be expressed as \cite{PhysRevLett.67.3231},
\begin{equation}
	\xi_g(\textbf{r}) =
	\begin{cases}
		0, & \text{if } r_{2n-1}< \textbf{r} \leq r_{2n}  \text{ (Transparent zone)}, \\
		1, & \text{if } r_{2n}< \textbf{r} \leq r_{2n+1}  \text{ (Opaque zone)}. \\
	\end{cases}   
	\label{geometrical_potential_energy_function} 
\end{equation}
Therefore, the geometrical potential energy of the FZP can be obtained as,
\begin{equation}
	V_g(\textbf{r})=V_{g0} \xi_g(\textbf{r}),
	\label{geometrical_potential_energy} 
\end{equation}
where $V_{g0}$ corresponds to the barrier height associated with the geometrical potential energy. The value of $V_{g0}$ is taken as $\sim 30E_{k0}$, which is significantly higher than the incident particle energy ($E_{k0}$ ) to minimize quantum tunneling through the opaque regions of the FZP \cite{PhysRevA.78.023612, 10.10635.0098030, barman2023near}.

\subsection{Atom-surface interaction energy ($V_{int}(\textbf{r})$)}
The atom-surface interaction potential energy ($V_{int}$), both inside and outside the zone plate, is governed by the long-range Casimir-van der Waals interaction. This interaction depends on the perpendicular distance (\( \chi \)) from the surface of the FZP, which can be expressed as \cite{PhysRevLett.86.987,Wang_2013,PhysRevA.95.043639, PhysRevA.65.032902}:
	\begin{equation}
		V_{int}(\chi) =-\frac{C_4}{(\chi+l)\chi^3},  
		\label{c_vw_potential} 
	\end{equation}
	where $l$ ($=9.3$ nm) is a characteristic length determining the transition from the van der Waals regime ($\propto 1/\chi^3$ at $\chi \ll l$) to the Casimir regime ($\propto 1/\chi^4$ at $\chi \gg l$) \cite{PhysRevLett.105.133203, PhysRevLett.91.193202, PhysRevA.101.022506}. $C_4$ ($=C_3 l$ with $C_3$ being the van der Waals adsorption coefficient) is the interaction coefficient of He atoms with Cr surface. The value of $C_3$ is taken as $2.5 \times 10^{-50}$ J m\textsuperscript{3} \cite{PhysRevA.78.010902}. Inside the zone plate, $\chi$ will be along $Y$ (see Fig. \ref{schematic_diagram_zone_plate}).

To avoid the singularity in Eq. \textcolor{blue}{(\ref{c_vw_potential})}, at $\chi=0$ (on the surface of FZP), a regularization of $V_{int}(\chi)$ is introduced, following a similar approach as outlined in Refs. \cite{Herwerth_2013, Galiffi_2017}. The interaction potential energy $V_{int}(\chi)$ is cut at $\chi=\Lambda$, and for $0 \leq \chi \leq \Lambda$, it is suitably continued as a parabolic curve with the vertex at $\chi=0$. All the parameters related to this regularization are obtained by imposing the continuity of $V_{int}(\chi)$ and $\partial V_{int} / \partial \chi$ at $\chi = \Lambda$, as proposed in Ref. \cite{Galiffi_2017}. The regularized form of $V_{int}(\chi)$ is obtained as,
\begin{widetext}
\begin{eqnarray}
	V_{int}(\chi) = \left\{
	\begin{array}{ll}
		-\frac{C_4}{(\chi+l)\chi^3}, & \textrm{if } \chi > \Lambda, \\
		\frac{C_4}{2\Lambda^5 (\Lambda +l)^2} \left[(4\Lambda + 3l) \chi^2 - (6\Lambda + 5l)\Lambda^2 \right], & \textrm{if } \Lambda \geq \chi > 0, \\
		-\frac{C_4}{2\Lambda^3} \frac{(6\Lambda + 5l)}{(\Lambda+l)^2}, & \textrm{if } \chi = 0. \\
	\end{array}
	\right.
	\label{eff_Casimir_van_der_Waals}
\end{eqnarray}
\end{widetext}
	
\begin{figure}
\centering
\includegraphics[scale=0.65]{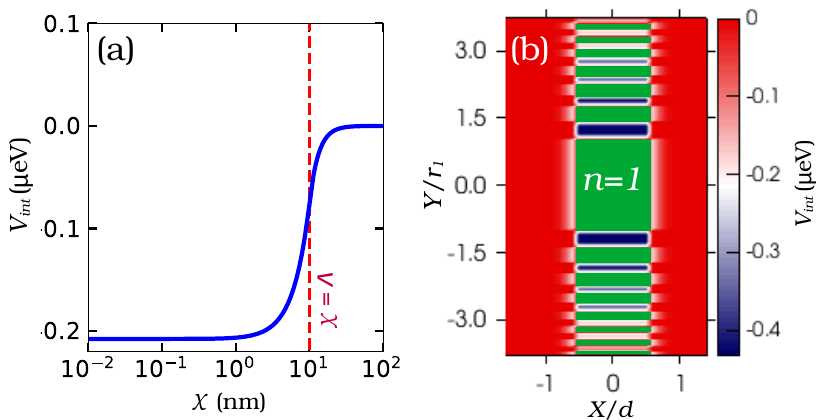}
\caption{(a) Variation of the regularized Casimir-van der Waals potential energy \( V_{int}(\chi) \) with the distance \( \chi \) from the surface of the Fresnel zone plate, as derived in Eq. \textcolor{blue}{(\ref{eff_Casimir_van_der_Waals})}. Here, the value of \( \Lambda \) is taken as 10 nm \cite{Galiffi_2017}. (b) Colormap of \( V_{int}(\chi) \) for both the inside and outside regions of the zone plate. The green color represents the opaque zones.}
\label{fig_casimir-van_der_waals_potential}
\end{figure}

Fig. \ref{fig_casimir-van_der_waals_potential}\textcolor{blue}{(a)} illustrates the variation of the regularized \( V_{int}(\chi) \) with respect to \( \chi \), as given by Eq. \textcolor{blue}{(\ref{eff_Casimir_van_der_Waals})}. It is evident that \( V_{int}(\chi) \) follows a parabolic curve within the region \( 0 \leq \chi \leq \Lambda \). In the simulations, the value of \(\Lambda\) is set to 10 nm \cite{Galiffi_2017}. The 2D colormap of \( V_{int}(\chi) \) for both the inside and outside regions of the zone plate is shown in Fig. \ref{fig_casimir-van_der_waals_potential}\textcolor{blue}{(b)}.

Retardation effects are included to accurately model the atom-surface interaction, capturing the transition from the van der Waals (\( \propto 1/\chi^3 \)) to the Casimir (\( \propto 1/\chi^4 \)) regimes, as also considered in an earlier study on slow atom diffraction \cite{PhysRevLett.127.170402}. These effects ensure consistency in the Casimir-Polder interaction relevant to the FZP geometry and the energy range studied, even if quantum reflection is not the primary focus. The atom-surface interaction model employed in this study balances computational efficiency and physical accuracy, making it well-suited for this case. The regularization ensures continuity at the surface, avoiding singularities and maintaining consistency with the Casimir-Polder interaction. While more advanced models \cite{https://doi.org/10.1002/andp.201500214, PhysRevA.94.023621} offer higher precision, their computational demands and detailed surface requirements limit practicality. Moreover, these models have been primarily employed in grating type of structures. This model effectively captures the essential physics of the interaction regime.

\begin{figure*}
	\centering
	\includegraphics[scale=0.8]{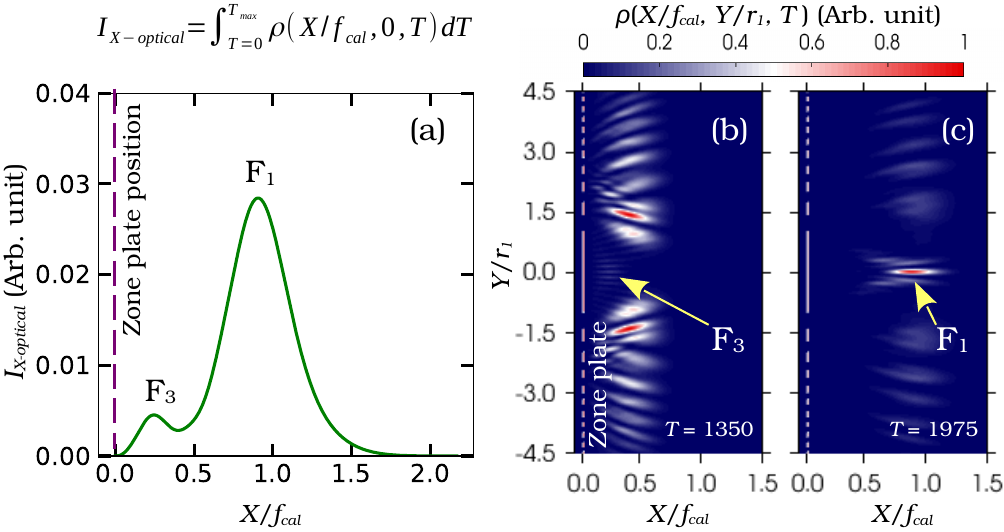}
	\caption{(a) Distribution of the integrated probability density $I_{X-\text{optical}}$, as defined in Eq. \textcolor{blue}{(\ref{I_optical_axis})}, for a Gaussian wave packet transmitted through a thin Fresnel zone plate with a thickness of $d=10$ nm along the optical axis ($X$-axis). The distribution of the probability density $\rho(X/f_{cal}, Y/r_1, T)$ of the transmitted wave packet is depicted at two specific time instances: (b) $T=1350$, corresponding to its arrival at the third focal point F\textsubscript{3}, and (c) $T=1975$, corresponding to its arrival at the first focal point F\textsubscript{1}. Here, the distributions of $\rho(X/f_{\text{cal}}, Y/r_1, T)$, shown in both the panels (b) and (c), have been normalized relative to the maximum value $\rho_{\text{max}}$ of the first-order peak.}
	\label{thin_grating_focusing}
\end{figure*}

\subsection{The induced polarization energy ($V_{induced}(\textbf{r})$)}

The FZP is biased with an electrostatic potential $V_n$, producing an electric field $\textbf{E}_b=-\bm{\textbf{}\nabla} V(\textbf{r})$, where $V(\textbf{r})$ is the spatial distribution of electric potential. Furthermore, in the presence of \( \mathbf{E}_{b} \), the He atom becomes polarized and acquires a dipole moment \( \mathbf{p} = \alpha \mathbf{E}_{b} \), where $\alpha$ ($=1.383746$ $a^3_0$ with $a_0$ being the Bohr radius) is the static polarizability of the He atom \cite{PhysRevA.68.012508}. When this polarized atom is in front of the metallic zone plate, it induces an image dipole \( \mathbf{p}_i \). According to the superposition principle, the total electric field experienced by the atom is the sum of the external field \( \mathbf{E}_{b} \) and the electric field \( \mathbf{E}_{i} \) due to its image dipole. Therefore, the induced potential energy experienced by the atom is given by \cite{PhysRevA.77.043406}:
\begin{align}
	V_{induced}(\textbf{r}) &= - \frac{\alpha}{2}|\textbf{E}_{b}+ \textbf{E}_{i}|^2, \nonumber \\ & =- \frac{\alpha}{2}\bigg[ |\textbf{E}_{b}|^2 + |\textbf{E}_{i}|^2 + 2 \textbf{E}_{b}\cdot \textbf{E}_{i} \bigg].
	\label{eqn_V_induced_full}
\end{align}

The first term on the right-hand side of Eq. \textcolor{blue}{(\ref{eqn_V_induced_full})} represents the Stark shift, which does not explicitly depend on \(\chi\). The second term can be neglected because it is of the order of \((\alpha \chi^{-3})^2 \), making it significantly smaller than the cross-term (third term), which is of the order of \(\alpha \chi^{-3}\), provided \((\alpha \chi^{-3} / 4 \pi \epsilon_0) \ll 1 \) \cite{PhysRevA.77.043406}. The cross-term can be obtained by first calculating \(\mathbf{E}_i\), given by \(\mathbf{E}_i = (1/4\pi \epsilon_0)[3(\mathbf{p}_i \cdot \hat X) \hat X - \mathbf{p}_i]/(2\chi)^3 \). Finally, the induced potential energy is obtained as,
\begin{align}
	V_{\text{induced}}(\mathbf{r}) &= -\frac{\alpha}{2} |\mathbf{E}_b|^2 -\frac{\alpha^2 |\mathbf{E}_b|^2}{32 \pi \epsilon_0} \bigg[ \frac{(\mathbf{\hat E}_b\cdot \hat X)^2 +1}{\chi^3}  \bigg].
\end{align}
where \( \mathbf{\hat{E}}_b \) and \( \hat{X} \) are unit vectors representing the directions of the electric field \( \mathbf{E}_b \) and the position vector \( X \), respectively.

Prior to determining the distribution of $V_{induced}(\textbf{r})$, the solution for $V(\textbf{r})$ is obtained numerically by solving Laplace's equation using the successive over-relaxation method \cite{press1992Numerical, anagnostopoulos2016Computational}, as detailed in Ref. \cite{10.10635.0098030}. Furthermore, it was calculated that the second term in the equation, which contains \( \alpha^2 \), is \( 10^{-6} \) times smaller than the first term, which contains \( \alpha \).

\begin{table}[h]
	\caption{\label{simulation_parameters} Values of the parameters used in the simulation.}
	\begin{ruledtabular}
		\begin{tabular}{lcc}
			\textrm{Parameters}  & \textrm{Values} \\
			\colrule
			Width of the initial wave packet ($\sigma_x$, $\sigma_y$) & (50 nm, 600 nm) \\
			Position of the initial wave packet ($X_0$, $Y_0$) & ($-f_{cal}/2$, $0$) \\
			de Broglie Wavelength ($\lambda_{dB}$) & 49.84 nm \\
			Velocity of the incident He atom ($\text{v}$) & 2 m/s \\
			Radius of the first zone ($r_1$) & 200 nm \\
			Focal length of the FZP ($f_{cal}$) & 803.21 nm \\
			Thickness of the FZP ($d$) & 50 nm \\
			Geometrical potential barrier height ($V_{g0}$) & 24.39 meV \\
			Scaling factor for energy ($V_0$) & 0.21 $\mu$eV \\
			Scaling factor for length ($\beta$) & 5 nm \\
			Scaling factor for time ($\tau$) & 3.152 ns \\
			Biasing voltage ($V_1$) & 0-30 V \\
			Radial modulation factor ($k_E$) & 0.01-0.4 \\
		\end{tabular}
	\end{ruledtabular}
\end{table}

\subsection{Numerical solution of TDSE}
The dynamics of the matter wave associated with a He atom interacting with a FZP can be obtained by solving the time-dependent Schrödinger equation (TDSE) in two-dimensional space ($x$, $y$). In order to enhance the computational efficiency in simulations, the spatial and temporal domains are transformed into arbitrary units of space and time as  $\textbf{r}(x,y)=\beta \textbf{R}(X, Y) $ and $t=\tau T$, where $\beta=\sqrt{ \hbar^2/2mV_0}$ and $\tau=2m\beta^2/\hbar$ are the scaling factors of length and time, respectively. The corresponding scaling factor for energy can be obtained as $V_0=\hbar^2/2m\beta^2$. Hence, the 2D Laplacian in the ($X$, $Y$) domain can be expressed as $\bm{\nabla}^2 = (1/\beta^2) \bm{\nabla}_{\beta}^2 = (1/\beta^2)(\partial^2/\partial X^2 + \partial^2/\partial Y^2)$. Similarly, the potential energies are scaled as $W_{g} (\textbf{R})=(1/V_0)V_{g}(\textbf{r})$, $W_{int} (\textbf{R})=(1/V_0)V_{int}(\textbf{r})$, and $W_{induced} (\textbf{R})=(1/V_0)V_{induced}(\textbf{r})$. Therefore, the Hamiltonian \textbf{H} in Eq. \textcolor{blue}{(\ref{hamiltonian})} can be expressed in the ($\textbf{R}$, $T$) space as,
\begin{align}
	\textbf{H}_{\beta} = - \bm{\nabla}^2_\beta + W_{g}(\textbf{R}) + W_{int}(\textbf{R}) +  W_{induced}(\textbf{R}).	\label{hamiltonian_modified}
\end{align}

The nondimensionalized form of the TDSE in arbitrary space (\textbf{R}) and time ($T$) can be obtained as,
\begin{align}
	i\frac{\partial \psi (\textbf{R},T)}{\partial T} = \textbf{H}_{\beta} \psi(\textbf{R},T),
	\label{sch_eq_modified}
\end{align}
where $i=\sqrt{-1}$ and $\psi(\textbf{R},T)$ is the complex wave function associated with the particle.

The TDSE (Eq. \textcolor{blue}{(\ref{sch_eq_modified})}) is solved numerically using the split-step Fourier method \cite{mclachlan_quispel_2002,WANG200517} for an initial wave function $\psi(\textbf{R},T=0)$. The time evolution of the wave function from $T$ to $T+\Delta T$ is carried out by \cite{mclachlan_quispel_2002,WANG200517}
\begin{align}
	\psi (\textbf{R}, & T+\Delta T) = \bigg[ \exp{\bigg(-\frac{i \mathbf{H_{pot}} \Delta T}{2}}\bigg) \nonumber\\ & \times \boldsymbol{\mathcal{F}}^{-1} \bigg\{\exp{\bigg(-i \mathbf{H_{kin}} \Delta T}\bigg) \nonumber\\ & \times \mathcal{F} \bigg( \exp{\bigg(-\frac{i\mathbf{H_{pot}} \Delta T}{2}\bigg)} \psi (\textbf{R},T)\bigg)  \bigg\} \bigg],
	\label{split_step_time_evolution}
\end{align}
where $\Delta T$ is the temporal step size. $\mathbf{H_{kin}}=- \bm{\nabla}^2_\beta$ and $\mathbf{H_{pot}}=W_{g}(\textbf{R}) + W_{int}(\textbf{R}) + W_{induced}(\textbf{R})$ are the kinetic and potential energy part of $\mathbf{H}_{\beta}$ ($=\mathbf{H_{kin}} + \mathbf{H_{pot}}$), respectively, in Eq. \textcolor{blue}{(\ref{hamiltonian_modified})}. $\mathcal{F}$ and ${\mathcal{F}}^{-1}$ denote forward and inverse Fourier Transforms, respectively. The evolution operator $\mathbf{H_{\beta}}$ ($= \mathbf{H_{kin}} + \mathbf{H_{pot}}$) is split using the second-order version of the Baker-Campbell-Hausdorff formula \cite{sanz2018numerical, mclachlan_quispel_2002, MUSLU2005581}. This splitting is performed with an error of $O(\Delta T ^3)$ to obtain the time evolution of $\psi(\textbf{R},T=0)$ \cite{mclachlan_quispel_2002}. The split-step Fourier method has been extensively used for solving the time-dependent Schrödinger equation in various physical contexts \cite{FEIT1982412, 10.1063/1.5127856, SEMENOVA2021110061, fractalfract7020188}. Recently, this method has been applied to investigate the propagation of matter waves, as demonstrated in Ref. \cite{PhysRevResearch.6.023165}, further highlighting its versatility and accuracy in modeling quantum dynamics.

The initial wave packet is taken as a two-dimensional Gaussian wave packet given by,
\begin{widetext}
	\begin{align}
		\psi(\textbf{R}, T=0) = N \exp{\bigg[ -\frac{1}{2}\bigg\{\frac{|(\textbf{R}-\textbf{R}_0) \cdot \hat{\textbf{X}}|^2}{\sigma_X^2}  +\frac{|(\textbf{R}-\textbf{R}_0)\cdot \hat{\textbf{Y}}|^2}{\sigma_Y^2}\bigg\} + i\textbf{K}_0 \cdot (\textbf{R}-\textbf{R}_0)\bigg]},
		\label{initial_wave_packet}
	\end{align}
\end{widetext}
where $N$ is the normalization constant and $\textbf{R}_0 (X_0,Y_0)$ is the center of the wave packet at $T=0$. $\sigma_X$ ($=\sigma_x/\beta$) and $\sigma_Y$ ($=\sigma_y/\beta$) are the width of the wave packet along $\hat {X}$ and $\hat {Y}$ directions, respectively. $\textbf{K}_0$ ($= \hat{X} 2\pi/\lambda^{'}_{dB}$) is the wave number of the initial wave packet moving along the $\hat{X}$ direction. The values of $\sigma_X$, $\sigma_Y$, and $\lambda^{'}_{dB}$ ($=\lambda_{dB}/\beta$) are normalized with respect to the spatial scaling factor $\beta$.

In the simulations, artificial reflections occur when the wave packet hits the boundary of the numerical box. Therefore, absorptive boundary conditions, as implemented in Refs. \cite{barman2023near, Herwerth_2013}, are employed to prevent unwanted reflections from the boundaries. A 2D region ($X$, $Y$) with square grids ($\Delta X= \Delta Y$) having dimensions $701\times 701$ is considered. Spatial and temporal scaling factors are chosen as $\beta=5$ nm and $\tau=3.152$ ns, respectively. All the simulation parameters are listed in Table \ref{simulation_parameters}. The numerical calculations, initiated with a normalized initial Gaussian wave packet (Eq. \textcolor{blue}{(\ref{initial_wave_packet})}), are implemented using the Python programming language. Standard modules such as NumPy and SciPy are employed in the calculations. Convergence in the simulation is ensured by maintaining the unit norm of the wave packet \(\psi(\mathbf{R}, T)\) at each time step. This preservation is crucial before the wave packet reaches the boundary of the simulation domain, which typically occurs later in the process as the transmitted wave packet is focused by the FZP. To further validate convergence, the spatial grid size and temporal step were each doubled in several instances of the simulation, and the results remained consistent.

In the simulations, the choice of a particle velocity of 2 m/s in this study was made deliberately to serve as a computational benchmark for exploring the detailed physics of matter-wave focusing in the low-energy regime. Although experimental studies typically involve higher velocities, the selected velocity lies within the regime where quantum reflection and significant phase modulation effects are dominant, as observed in low-velocity quantum reflection studies. This regime allows for the investigation of subtle interaction effects, such as those arising from atom-surface potentials and external biasing fields, which may be less prominent at higher velocities.
	
Additionally, the simulations provide insights into focusing mechanisms that could be applicable to ultra-cold atomic beams or systems where velocity control is achievable through advanced techniques. While the lowest experimentally achieved velocities have been reported at 20 m/s \cite{PhysRevLett.127.170402}, the use of dilution refrigeration could attain velocities corresponding to temperatures as low as ~1 mK \cite{PhysicsPhysiqueFizika.4.1, Pobell2007}. However, it is important to note that helium beam intensities under such conditions would likely be significantly reduced. This study extends the exploration into a theoretical regime to establish a understanding that may guide future experimental advancements and look at effects as mentioned above arising at low velocities. The model can be extended to accommodate higher velocities across a broader range.

\begin{figure*}
\centering
\includegraphics[scale=0.83]{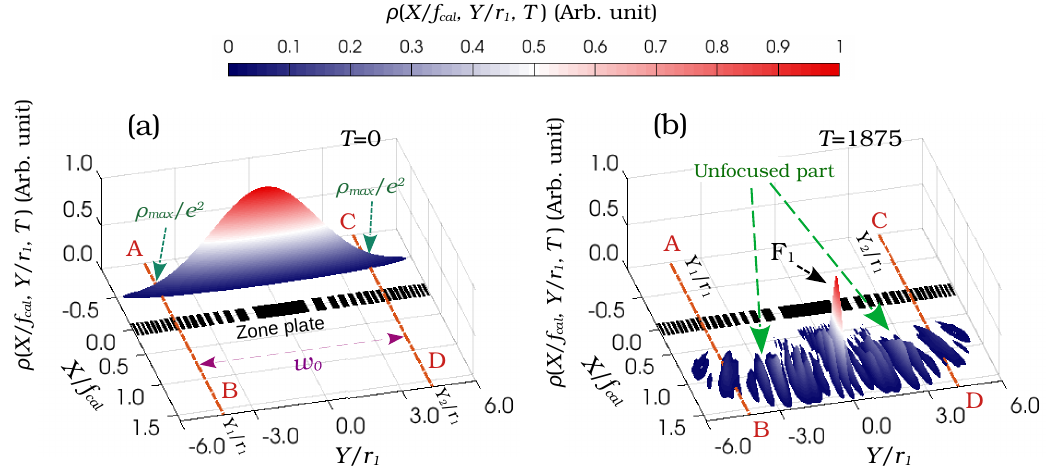}
\caption{(a) Distribution of the probability density $\rho(X/f_{cal}, Y/r_1, 0)=|\psi(X/f_{cal}, Y/r_1, 0)|^2$ of the initial Gaussian wave packet moving along the $+X$ direction. The width $w_0$ of the initial wave packet is defined at the waist (AB and CD dotted lines), where the probability density $\rho(X/f_{cal}, Y/r_1, 0)$ is reduced to $1/e^2$ ($\sim 13.5\%$) of its maximum value. (b) Distribution of $\rho(X/f_{cal}, Y/r_1, T)=|\psi(X/f_{cal}, Y/r_1, T)|^2$ of the transmitted wave packet at  $T_1=1875$, when the wave packet reached at the focal point F\textsubscript{1} ($X=f_1$). Here, the reflected part is absorbed at the boundary of the simulation domain.} \label{figure_initial_wave_packet}
\end{figure*}

\section{Results and discussions} \label{result_and_discussions}


\subsection{Benchmarking of the simulation}\label{benchmarking}

To validate the simulations, an initial investigation of a thin Fresnel zone plate (thickness $d\sim 10$ nm) was conducted. The obtained results were subsequently compared with analytically derived findings reported earlier \cite{10.1119/1.18587, Greve:13}. For this analysis, a straightforward scenario was considered, excluding the Casimir-van der Waals interaction potential energy $V_{int}(\text{\textbf{r}})$ and without the application of any electrical bias.

In the simulation, an initial Gaussian wave packet was considered, as defined by Eq. \textcolor{blue}{(\ref{initial_wave_packet})}. The wave packet undergoes temporal evolution and gets diffracted by the FZP as it propagates along the $+X$ direction. Upon interaction with the FZP, the wave packet experiences partial reflection and transmission. The transmission coefficient $T_c$ for the FZP was calculated by,
\begin{align}
	T_c =\int_{Y=-\infty}^{+\infty} dY \int_{X=0}^{+\infty} dX |\psi(\textbf{R},T)|^2.
	\label{Tc_equn}
\end{align}

A transmission coefficient of $T_c = 0.358$ has been obtained for the thin FZP. A similar value of the transmission coefficient has been reported earlier for a standard FZP in Ref. \cite{Greve:13}. 

To determine the positions of the focal points, the probability density along the optical axis $OX$ (see Fig. \ref{schematic_diagram_zone_plate}) was recorded at each time step \cite{deng2017carbon}. Subsequently, a time-integrated intensity profile along $OX$ was obtained as,
\begin{align}
	I_{X-optical} (X, Y=0) = \int_{T=0}^{T_{max}} |\psi(X, Y=0,T)|^2 dT,
	\label{I_optical_axis}
\end{align}
where $T_{max}$ is the total simulation time.

The simulation result, illustrating the distribution of $I_{X-optical}$ along the $X$-axis for the thin FZP, is shown in Fig. \ref{thin_grating_focusing}\textcolor{blue}{(a)}. Multi-focusing behavior is observed in this case, with distinct focal points, namely F\textsubscript{1} and F\textsubscript{3}, as depicted in Fig. \ref{thin_grating_focusing}\textcolor{blue}{(a)}. Furthermore, it is observed that the first focal length $f_1$ is approximately equal to the calculated focal length, $f_{\text{cal}}$ ($= r^2_1/\lambda_{dB}$). Additionally, the focal length for F\textsubscript{3} is found to be $f_3 \approx f_1/3$. This confirms the odd-order multi-focusing by the FZP, in agreement with the findings reported earlier \cite{10.1119/1.18587}. Moreover, the intensity at F\textsubscript{3} is observed to be significantly lower ($\sim 16\%$) in comparison to the intensity at F\textsubscript{1}, as shown in Fig. \ref{thin_grating_focusing}\textcolor{blue}{(a)}. This aspect is further clarified by analyzing the probability density distributions at both focal points (F\textsubscript{3} and F\textsubscript{1}), as shown in Fig. \ref{thin_grating_focusing}\textcolor{blue}{(b)} and \textcolor{blue}{(c)}, respectively. 

Thereafter, the focusing efficiency $\eta_j$ at the $j$-th order focal point was calculated as \cite{10.1119/1.18587}, 

\begin{align}
	\eta_j=\frac{I_j}{I_{initial}},
	\label{efficiency}   
\end{align}
where, $I_{initial}$ and $I_j$ are the intensity of the incident wave packet and the intensity at $j$-th focal point. In this case, the values of $\eta_1$ and $\eta_3$ are obtained as $9.92\%$ and $0.84\%$, respectively. A similar value of $\eta_1$ ( $= 1/\pi^2 \approx 10\%$) has been previously reported for a standard FZP \cite{10.1119/1.18587, Greve:13}. These observations validate the accuracy of the simulation. In all subsequent cases, the actual system, including all interaction potential energies, has been simulated.
\subsection{Diffractive focusing of matter waves by FZP}\label{zone_plate_focusing}

In this case, the diffractive focusing by a metallic zone plate with a thickness of $d=50$ nm has been investigated. All the other simulation parameters are listed in Table \ref{simulation_parameters}. The distribution of the probability density $\rho(X, Y, T=0)$ of the initial wave packet is shown in Fig. \ref{figure_initial_wave_packet}\textcolor{blue}{(a)}. The width ($w_0$) of the initial wave packet is determined at the waist of the Gaussian distribution, as indicated by the dotted lines AB (at $Y/r_1=Y_1/r_1=-4.225$) and CD (at $Y/r_1=Y_2/r_1=4.225$) in Fig. \ref{figure_initial_wave_packet}\textcolor{blue}{(a)}. At the waist, the probability density $\rho(X, Y, 0)$ diminishes to $1/e^2$ ($\sim 13.5\%$) of its peak value ($\rho_{max} = 2.627 \times 10^{-4}$) \cite{PhysRevLett.88.100404, Self:83}.

In this case, a transmission of $23.3\%$ is observed for the wave packet through the FZP, which is relatively smaller than the transmission observed in a thin grating (35.8\%). This is due to the increased thickness of the FZP. The spatial distribution of the transmitted wave packet $\rho(X, Y, T)$ after the FZP at time $T=1875$ is shown in Fig. \ref{figure_initial_wave_packet}\textcolor{blue}{(b)}. The wave packet is observed to be focused at the first focal point (F\textsubscript{1}) with a focal length $f_1$ ($\approx 675$ nm). Notably, it is observed that the focal length $f_1$ is approximately $84.3\%$ of the calculated focal length $f_{\text{cal}}$ ($\sim 800$ nm). The reduction in the focal length is attributed to the finite width of the zone plate ($d=50$ nm) employed in this study.

In Fig. \ref{figure_initial_wave_packet}\textcolor{blue}{(b)}, it can be observed that, in addition to the central peak at F\textsubscript{1}, multiple off-axis peaks are present with reduced intensity. These peaks, between the dotted lines AB and CD in Fig. \ref{figure_initial_wave_packet}\textcolor{blue}{(b)}, primarily originate from zeroth-order diffraction, representing the unfocused part of the wave packet \cite{Eder_2012,PhysRevA.95.023618}. The presence of these unfocused parts of the wave packet decreases the quality of focusing.

\begin{figure}[h]
\centering
\includegraphics[scale=0.85]{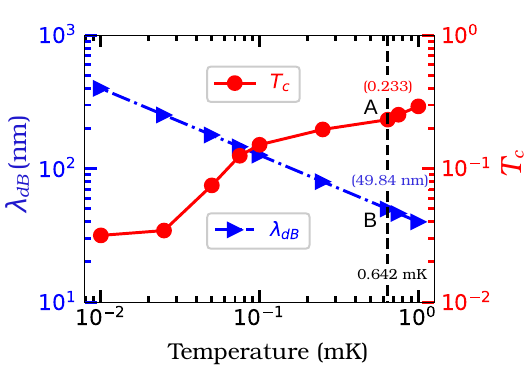}
	\caption{Variation of the de Broglie wavelength ($\lambda_{dB}$) of He atom with temperature $T_a$ is shown on the left $y$-axis. The variation of the transmission coefficient ($T_c$) with temperature, as it passes through the Fresnel zone plate, is shown on the right $y$-axis.}
\label{transmission_vs_temperature}
\end{figure}

\begin{figure*}
	\centering
	\includegraphics[scale=0.8]{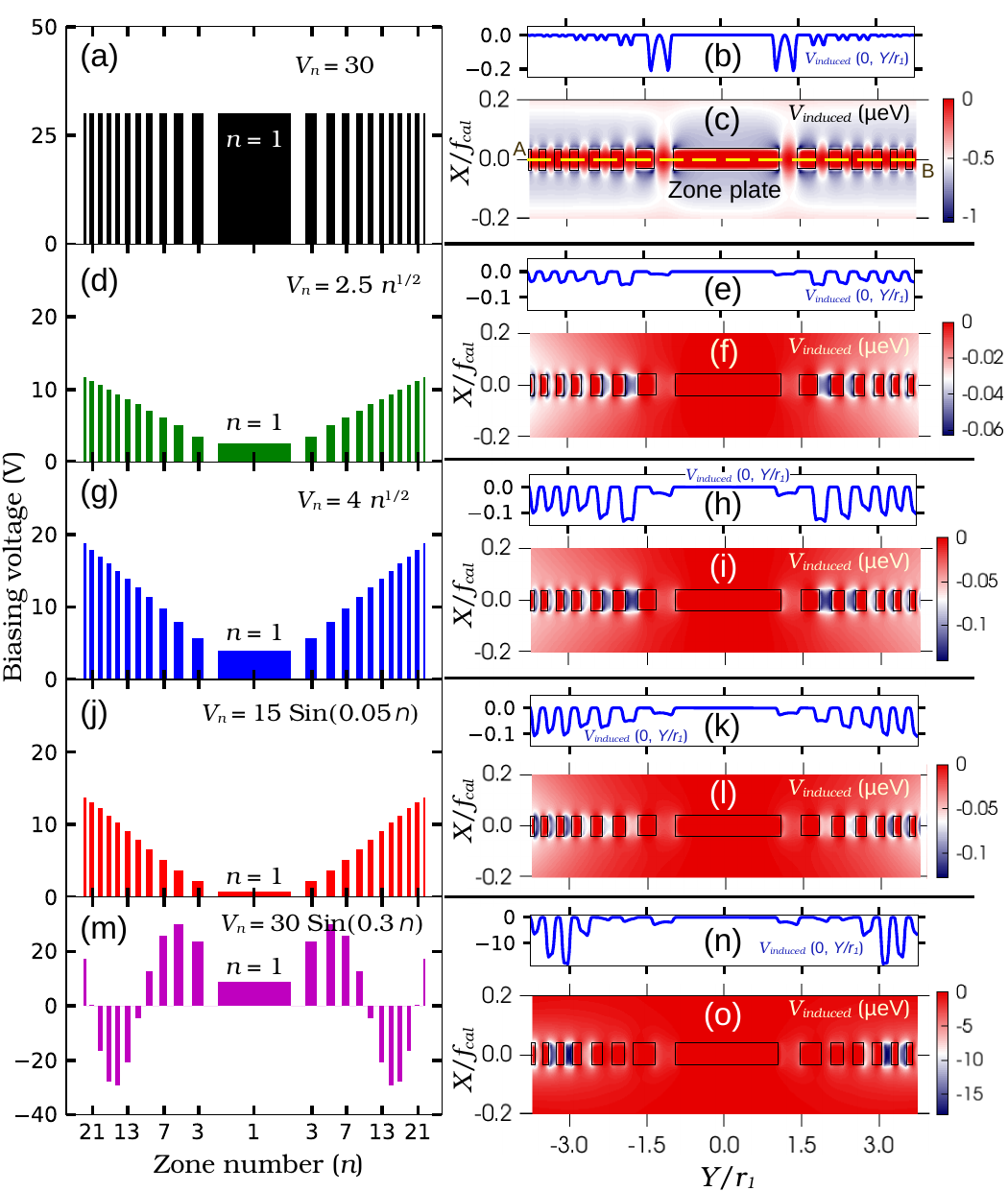}
	\caption{Distributions of the biasing potentials applied to the $n$th opaque zone are illustrated for different scenarios: (a) $V_n = 30$ V, (d) $V_n = 2.5 \sqrt{n}$ V, (g) $V_n = 4 \sqrt{n}$ V, (j) $V_n = 15 \sin(0.05 n)$ V, and (m) $V_n = 30 \sin(0.3 n)$ V. The corresponding distributions of the induced potential $V_{\text{induced}}(X/f_{\text{cal}}, Y/r_1)$ arising from the polarization of a helium atom in the presence of electric fields are shown in panels (c), (f), (i), (l), and (o), respectively. The panels (b), (e), (h), (k), and (n) depict the variation of $V_{\text{induced}}(0, Y/r_1)$ in each case, respectively.}
	\label{biasing_voltage}
\end{figure*}
\subsection{Effect of particle temperature on matter wave focusing} \label{temperature_dependence}
Simulations were conducted at various incident particle temperatures \( T_a \) to investigate their influence on the transmission of the wave packet through the FZP, primarily characterized by its propagation velocity. The de Broglie wavelength $\lambda_{dB}$ varies with $T_a$ as,
\begin{align}
	\lambda_{dB} =\frac{h}{\sqrt{3mk_B T_a}},
	\label{lambda_dB_vs_temp}
\end{align}
where $k_B$ is the Boltzmann constant.

The variation of $\lambda_{dB}$ with $T_a$ is shown on the left y-axis of Fig. \ref{transmission_vs_temperature}. From the figure, it is evident that $\lambda_{dB}$ decreases monotonically as $T_a$ increases, as expected. The right y-axis of Fig. \ref{transmission_vs_temperature} depicts the variation of the $T_c$ with $T_a$. It is observed that the transmission coefficient \( T_c \) increases with the increase in \( T_a \). The transmission coefficient begins to saturate at \( T_a = 0.1 \) mK. For subsequent simulations, a constant temperature of 0.642 mK is used, which corresponds to a de Broglie wavelength \( \lambda_{dB} = 49.84 \) nm, at which the transmission coefficient \( T_c \) is \( 0.233 \).

\subsection{Electrostatic biasing of FZP}\label{electrostatic_biasing}
The focusing properties of the FZP are investigated using three distinct biasing configurations: (i) uniform biasing of all zones with a constant potential, \( V_n = V_1 \), where \( V_1 \) represents the applied voltage to the central zone; (ii) biasing of the \(n\)th zone with \( V_n = V_1 \sqrt{n} \); and (iii) radially modulated voltage for the \(n\)th zone given by \( V_n = V_1 \sin(k_E n) \), where \( k_E \) is the modulation factor. These voltage configurations ($V_n$) are chosen to manipulate and control the phase differences among the wave fronts emerging from the different zones, which in turn affect the focusing properties. The phase shift due to the induced potential is given by \cite{RevModPhys.81.1051}:
\begin{align}
\Delta_{induced} = - \frac{1}{\hbar v} \int_d V_{induced}(\mathbf{r}) \, dl,
\end{align}
where \( \Delta_{induced} \) can be controlled by varying the biasing voltage configuration $V_n$. The results of different biasing configurations are discussed below.

\subsubsection{$V_n=V_1$}
In this case, the biasing voltage \(V_1\) varies from 0 to 30 V. Fig. \textcolor{blue}{\ref{biasing_voltage}(a)} illustrates the distribution of the biasing voltage $V_n$ across all the opaque zones for \(V_1 = 30\) V.

The distribution of \(V_{\text{induced}}(\textbf{r})\) for \(V_1 = 30\) V is shown in Fig. \textcolor{blue}{\ref{biasing_voltage}(c)}. The profile of \(V_{\text{induced}}(0, Y)\) along the FZP, as indicated by the yellow dotted line AB in Fig. \textcolor{blue}{\ref{biasing_voltage}(c)}, is depicted in Fig. \textcolor{blue}{\ref{biasing_voltage}(b)}. From Fig. \textcolor{blue}{\ref{biasing_voltage}(c)}, it is evident that \(V_{\text{induced}}(0, Y)\) is nearly zero inside the transparent zones, whereas it has an enhanced magnitude near the edges of the opaque zones. This indicates that the effective width (\(= r_n - r_{n-1}\)) of different zones remains unchanged after applying a biasing voltage to the FZP as \(V_n = V_1\). Instead, it is seen in Fig. \textcolor{blue}{\ref{biasing_voltage}(c)} that the biasing in this manner effectively alters the cross-sectional shape of the zones from rectangular to circular, as previously reported for a diffraction grating \cite{10.10635.0098030}. Furthermore, the induced potential energy gradually decreases with the normal distance \(\chi\) away from the zone plate (along the \(X\)-axis), as illustrated in Fig. \textcolor{blue}{\ref{biasing_voltage}(c)}. 

\subsubsection{$V_n = V_1 \sqrt{n}$}
In this case, the value of \(V_1\) is varied from 0 to 5 V. The variation of \(V_n\) is shown in Figs. \textcolor{blue}{\ref{biasing_voltage}(d)} and \textcolor{blue}{\ref{biasing_voltage}(g)} for \(V_1 = 2.5\) V and 4 V, respectively.

Since $V_{\text{induced}}(\mathbf{r})$ is attractive in nature, the transmission of the wave packet through the transparent zones will be proportional to the magnitude of the induced potential inside the transparent zones. The variation of the induced potential energy inside the transparent zones along the $Y$-axis can be obtained as,

\begin{align}
	V_{induced} (0, Y) = - \frac{\alpha}{2} \bigg( - \frac{d V_n}{d n} \frac{d n}{d r}\bigg)^2 = - \frac{\alpha V_1^2}{2\lambda_{dB} f_{cal}},
	\label{V_induced_along_zone_plate_root_N}
\end{align}
where $dY$ is replaced by $dr$ because $dY=dr$. It can be noticed that $V_{\text{induced}}(0, Y)$ is independent of $n$ and varies quadratically with $V_1^2$. Therefore, the value of $V_{\text{induced}}(0, Y)$ will be the same inside all the transparent zones. Moreover, it can be expected that the transmission must vary quadratically with $V_1$.

The distributions of $V_{induced}(\text{\textbf{r}})$ for $V_1=2.5$ V and $4$ V are illustrated in Fig. \textcolor{blue}{\ref{biasing_voltage}(f)} and \textcolor{blue}{\ref{biasing_voltage}(i)}, respectively. It can be seen that larger values of $V_{\text{induced}}(\textbf{r})$ are attained inside the transparent zones compared to the exterior of the FZP. This is also evident from Figs. \textcolor{blue}{\ref{biasing_voltage}(e)} and \textcolor{blue}{(h)}, where the distributions of $V_{\text{induced}} (0, Y)$ are shown for $V_1=2.5$ V and $4$ V, respectively. In Figs. \textcolor{blue}{\ref{biasing_voltage}(e)} and \textcolor{blue}{(h)}, it can be seen that the magnitude of $V_{\text{induced}} (0, Y)$ is almost constant inside the higher order transperent zones. Moreover, the cross-sectional shape of the opaque zones is altered by this, which differs from the effects observed in the first case ($V_n=V_1$). Therefore, by biasing the FZP in this manner, the transmission of the wave packet can be modulated differently than in the first case, thereby allowing for controlled manipulation of the focusing properties of the transmitted wave packet.

\subsubsection{$V_n = V_1 \sin(k_E n)$}
In this case, \(k_E\) is an additional controlling factor that offers increased flexibility in manipulating matter waves by the FZP. The values of \(V_1\) are varied from 0 to 30 V, and the values of \(k_E\) are varied from 0.01 to 0.4.

The variation of $V_{\text{induced}}(0, Y)$ can be derived for this case as,

\begin{align}
	V_{induced} (0, Y) = - \frac{2 \alpha V_1^2 k_E^2}{\lambda_{dB} f_{cal}} n \cos^2(k_E n).
	\label{V_induced_along_zone_plate_sin_kN}
\end{align}

Here, it can be observed that $V_{\text{induced}}(0, Y)$ varies sinusoidally with both $k_E$ and $n$. Therefore, it is expected that the transmission coefficient will also vary in a similar manner as $V_{\text{induced}}(0, Y)$ with $k_E$ and $n$.

Figs. \textcolor{blue}{\ref{biasing_voltage}(j)} and \textcolor{blue}{(m)} depict the distribution of the biasing voltage to the opaque zones for ($V_1=15$ V, $k_E=0.05$) and ($V_1=30$ V, $k_E=0.3$), respectively. It is observed that, for higher values of $k_E$ ($=0.3$), the sinusoidal modulation in the biasing voltage applied to different zones becomes more prominent (see Fig. \textcolor{blue}{\ref{biasing_voltage}(m)}), particularly near the central part of the FZP, where the wave packet primarily interacts.

The distributions of $V_{\text{induced}}(\text{\textbf{r}})$ for ($V_1=15$ V, $k_E=0.05$) and ($V_1=30$ V, $k_E=0.3$) are shown in Figs. \textcolor{blue}{\ref{biasing_voltage}(l)} and \textcolor{blue}{(o)}, respectively. In Fig. \textcolor{blue}{\ref{biasing_voltage}(o)}, it can be observed that $V_{\text{induced}}(\text{\textbf{r}})$ is notably small in the proximity of the central zone ($n=1$) and undergoes modulation as one moves towards the higher-order zones. Additionally, a decrease in $V_{\text{induced}}(\textbf{r})$ is noted in the exterior region compared to inside the transparent zones. Figs. \textcolor{blue}{\ref{biasing_voltage}(k)} and \textcolor{blue}{(n)} illustrate the distributions of $V_{\text{induced}} (0, Y)$ for ($V_1=15$ V, $k_E=0.05$) and ($V_1=30$ V, $k_E=0.3$), respectively. In Fig. \textcolor{blue}{\ref{biasing_voltage}(n)}, it is observed that the variation of the amplitude of $V_{\text{induced}}(0, Y)$ inside different zones for ($V_1=30$ V, $k_E=0.3$) agrees well with the results obtained using Eq. \textcolor{blue}{(\ref{V_induced_along_zone_plate_sin_kN})}. For smaller values of $k\leq 0.05$, it can be shown that $V_{\text{induced}} (0, Y) = - (\alpha V_1^2 k_E^2/\lambda_{dB} f_{\text{cal}}) n$, implying a linear dependence on $n$. This is also evident from Fig. \textcolor{blue}{\ref{biasing_voltage}(k)}.

Moreover, a periodic modulation in the cross-sectional shape of the opaque zones is also evident from Fig. \textcolor{blue}{\ref{biasing_voltage}(o)}, in contrast to the other two cases mentioned earlier. Therefore, biasing the FZP sinusoidally can lead to a different modulation of the diffraction of the wave packet compared to the other two cases.

\subsection{Focusing properties of a biased FZP}\label{electrostatic_biasing}

Here, the effects of biasing the FZP on the distribution of the transmitted wave packet $\rho(X, Y, T)$, transmission coefficient ($T_c$), and focusing properties, such as focal length ($f$), size of the focused wave packet ($\sigma_f$) and focusing efficiency ($\eta$), are demonstrated.

\begin{figure}
	\centering
	\includegraphics[scale=0.9]{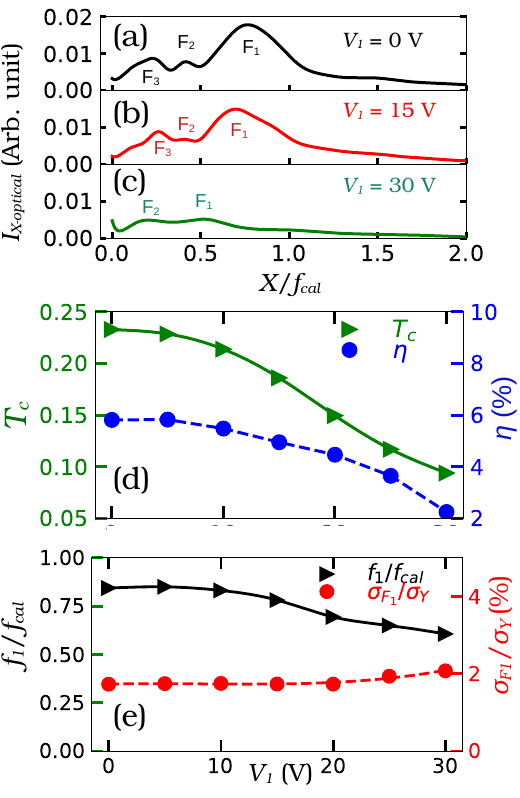}
	\caption{Distributions of the integrated probability density \(I_{X-\text{optical}}\) along the optical axis (\(X\)-axis) when the \(n\)th zone of the Fresnel zone plate (FZP) is biased as \(V_n = V_1\), for different biasing voltages: (a) \(V_1 = 0\) V, (b) \(V_1 = 15\) V, and (c) \(V_1 = 30\) V. Panel (d) shows the variation of the transmission coefficient \(T_c\) (left \(y\)-axis) and focusing efficiency \(\eta_1\) (right \(y\)-axis). Panel (e) shows the variation of the first focal length \(f_1\) (left \(y\)-axis) and the width (\(\sigma_{F1}\)) of the focused wave packet at the first focal point (right \(y\)-axis).}
	\label{Normal_Case_axial_I_Tc_Eta}
\end{figure}

\subsubsection{$V_n=V_1$}
To determine the focal points, the distributions of the integrated probability density \(I_{X-\text{optical}}\) (Eq. \textcolor{blue}{(\ref{I_optical_axis})}) along the optical axis (\(X\)-axis) are obtained for different values of \(V_1\) and shown in Figs. \ref{Normal_Case_axial_I_Tc_Eta}\textcolor{blue}{(a)}-\textcolor{blue}{(c)}. In Fig. \ref{Normal_Case_axial_I_Tc_Eta}\textcolor{blue}{(a)}, where the distribution of \(I_{X-\text{optical}}\) is shown for \(V_1 = 0\), three focal points (F\textsubscript{1}, F\textsubscript{2}, and F\textsubscript{3}) are observed. However, the intensity at the higher-order focal points is significantly lower than at F\textsubscript{1}. Additionally, the first focal length \(f_1\) is found to be approximately 84.3\% of the calculated focal length \(f_{\text{cal}} (= r_1^2 / \lambda_{\text{dB}})\) due to the finite width \((d = 50\) nm) of the FZP, a factor typically not considered in analytical calculations. More importantly, it can be observed that as \(V_1\) increases from 0 V to 30 V, the intensity at \(F_1\) decreases, as depicted in Figs. \ref{Normal_Case_axial_I_Tc_Eta}\textcolor{blue}{(a)}-\textcolor{blue}{(c)}.

Thereafter, the transmission coefficient \(T_c\) of the wave packet through the FZP has been obtained using Eq. \textcolor{blue}{(\ref{Tc_equn})}. The left \(y\)-axis of Fig. \ref{Normal_Case_axial_I_Tc_Eta}\textcolor{blue}{(d)} shows the variation of \(T_c\) with the biasing voltage \(V_1\). It is observed that \(T_c\) decreases from 0.233 to 0.094 as \(V_1\) increases from 0 to 30 V. This occurs because the induced polarization energy \(V_{\text{induced}}(\mathbf{r})\) increases on both sides of the FZP with the increase in biasing voltage \(V_1\), as depicted in Fig. \textcolor{blue}{\ref{biasing_voltage}(c)}. Hence, \(V_{\text{induced}}(\mathbf{r})\) begins to dominate the long-range atom-surface interaction \(V_{\text{int}}(\mathbf{r})\) (see Fig. \ref{fig_casimir-van_der_waals_potential}), resulting in the modulation of the overall interaction between the He atom and the FZP. This modulation enhances the probability of quantum reflection of He atoms by the FZP, consequently leading to a decrease in the transmission probability with the increase in \(V_1\).

The focusing efficiency at the \(j\)th order focal point F\textsubscript{j} has been calculated using Eq. \textcolor{blue}{(\ref{efficiency})}. The variation of \(\eta_1\) with \(V_1\) is shown on the right \(y\)-axis of Fig. \ref{Normal_Case_axial_I_Tc_Eta}\textcolor{blue}{(d)}. It is observed that, similar to the transmission coefficient, \(\eta_1\) also decreases from 5.82\% to 2.25\% with the increase in the biasing voltage \(V_1\). Moreover, the values of \(\eta_j\) at higher-order (\(j \geq 2\)) focal points are found to vary under 2.5\% for all values of \(V_1\) (not shown).

The variation of the first focal length \(f_1\) with \(V_1\) is shown on the left \(y\)-axis of Fig. \ref{Normal_Case_axial_I_Tc_Eta}\textcolor{blue}{(e)}. It is observed that \(f_1\) decreases from \(0.84f_{\text{cal}}\) to \(0.61f_{\text{cal}}\) as the value of \(V_1\) increases from 0 to 30 V. However, the width \(\sigma_{F1}\) (full width at half maximum) of the focused wave packet at F\textsubscript{1} remains almost constant at 1.85\% of the width of the initial wave packet \(\sigma_Y\), as illustrated on the right \(y\)-axis in Fig. \ref{Normal_Case_axial_I_Tc_Eta}\textcolor{blue}{(e)}.

\begin{figure}
\centering
\includegraphics[scale=0.9]{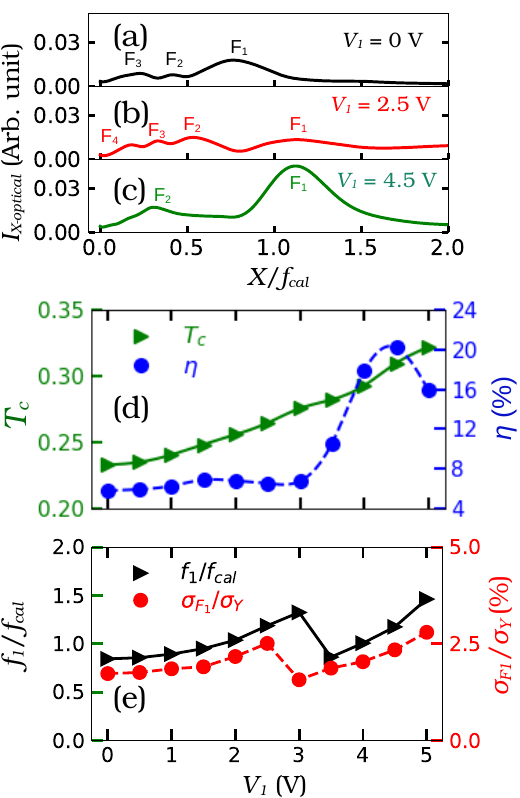}
\caption{Distributions of the integrated probability density \(I_{X-\text{optical}}\) along the optical axis (\(X\)-axis) when the \(n\)th zone of the Fresnel zone plate (FZP) is biased as \(V_n = V_1 \sqrt{n}\), for different biasing voltages: (a) \(V_1 = 0\) V, (b) \(V_1 = 2.5\) V, and (c) \(V_1 = 4.5\) V. Panel (d) shows the variation of the transmission coefficient \(T_c\) (left \(y\)-axis) and focusing efficiency \(\eta_1\) (right \(y\)-axis). Panel (e) shows the variation of the first focal length \(f_1\) (left \(y\)-axis) and the width (\(\sigma_{F1}\)) of the focused wave packet at the first focal point (right \(y\)-axis).}
\label{axial_I_Tc_Eta_root_n}
\end{figure}

\begin{figure*}
	\centering
	\includegraphics[scale=0.8]{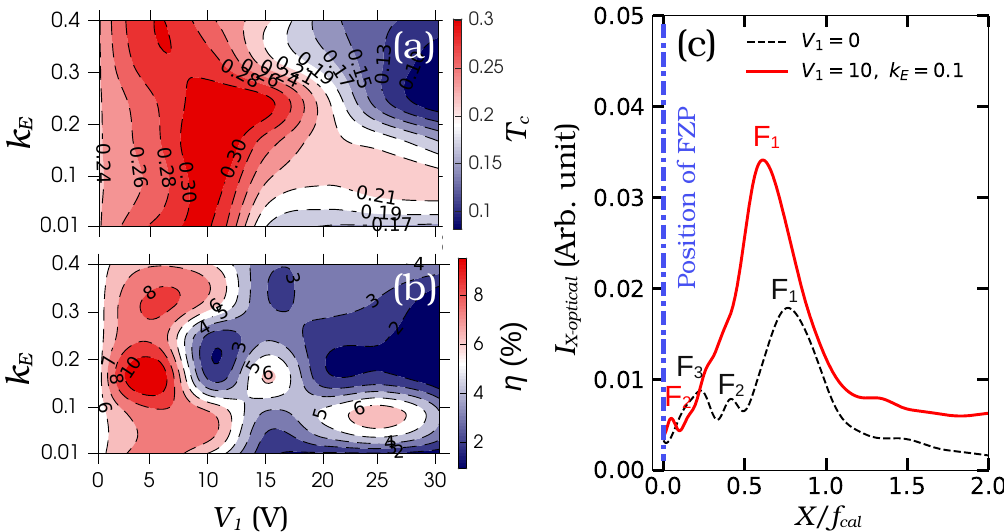}
	\caption{Variations of (a) the transmission coefficient \( T_c \) and (b) the efficiency \( \eta_1 \) for first-order focusing with the bias voltage \( V_1 \) and the radial modulation factor \( k_E \). Here, the \( n \)th zone of the Fresnel zone plate (FZP) is biased as \( V_n = V_1 \sin(k_E n) \). (c) Variation of the time-integrated intensity along the optical axis \( I_{X-\text{optical}} \) is shown for \( (V_1 = 0, k_E = 0) \) (black dotted curve) and for \( (V_1 = 10, k_E = 0.1) \) (red solid curve).}
	\label{Tc_Eta_Axial_profile_Sin_kN}
\end{figure*}

\subsubsection{$V_n=V_1\sqrt{n}$}
The positions of the focal points at different biasing voltages \(V_1\) are shown in Figs. \ref{axial_I_Tc_Eta_root_n}\textcolor{blue}{(a)}-\textcolor{blue}{(c)}, where the distributions of \(I_{X-\text{optical}}\) along the optical axis are presented. In Figs. \ref{axial_I_Tc_Eta_root_n}\textcolor{blue}{(a)}-\textcolor{blue}{(c)}, it can be observed that the number of focal points depends on the biasing voltage. More interestingly, it can be seen that the peak intensity at F\textsubscript{1} for \(V_1 = 4.5\) V is larger compared to the other cases.

The variation of the transmission coefficient $T_c$ is shown on the left $y$-axis of Fig. \ref{axial_I_Tc_Eta_root_n}\textcolor{blue}{(d)}. It is observed that $T_c$ varies quadratically with $V_1$, as mentioned earlier, and increases from 0.233 to 0.322 as $V_1$ increases from 0 to 5 V. This can be understood through the distribution of $V_{\text{induced}}(\mathbf{r})$ resulting from the biasing of the FZP as $V_n=V_1\sqrt{n}$. In Figs. \ref{biasing_voltage}\textcolor{blue}{(f)} and \textcolor{blue}{(i)}, it is observed that the magnitude of $V_{\text{induced}}(\mathbf{r})$ is nearly smaller near the central zone and it increases inside the higher-order transparent zones. This effectively enlarges the width of the higher-order transparent zones, thereby increasing their effective area. Hence, there is an increase in wave packet transmission with increase in $V_1$. Furthermore, unlike the previous case ($V_n=V_1$), $V_{\text{induced}}(\mathbf{r})$ has a small magnitude on either side of the FZP (see Figs. \ref{biasing_voltage}\textcolor{blue}{(f)} and \textcolor{blue}{(i)}) due to the application of a relatively smaller biasing voltage $V_1$ (maximum voltage 5 V) to the central zone. Hence, the reflection probability is not enhanced as observed in the previous case. 

The variation of \(\eta_1\) with \(V_1\) is shown on the right \(y\)-axis of Fig. \ref{axial_I_Tc_Eta_root_n}\textcolor{blue}{(d)}. It is observed that the value of \(\eta_1\) strongly depends on \(V_1\). Initially, \(\eta_1\) remains almost constant at approximately 6\% for \(V_1 \leq 3.0\) V. Thereafter, with the increase in \(V_1\), \(\eta_1\) increases, attaining a maximum value of 20.2\% at \(V_1 = 4.5\) V, and then decreases with further increases in \(V_1\), as shown in Fig. \ref{axial_I_Tc_Eta_root_n}\textcolor{blue}{(d)}. Moreover, it is observed that the focusing efficiency at higher-order focal points (\(\eta_j\), \(j \geq 2\)) also depends on \(V_1\), but it remains within 5\% (not shown).

The variation of \(f_1\) with \(V_1\) is shown in Fig. \ref{axial_I_Tc_Eta_root_n}\textcolor{blue}{(e)} (left \(y\)-axis). It can be seen that \(f_1\) increases from \(0.843f_{\text{cal}}\) to \(1.463 f_{\text{cal}}\) as \(V_1\) increases from 0 to 5 V. This results from the increase in the effective radius of the transparent zones with the increase in \(V_1\), as depicted in Figs. \ref{biasing_voltage}\textcolor{blue}{(f)} and \textcolor{blue}{(i)}. As the first focal length \(f_1\) is proportional to \(r^2_n/n\), in contrast to the previous case, the focal length \(f_1\) decreases with the increase in \(V_1\).

The variation of \(\sigma_{F1}\) with \(V_1\) is shown in Fig. \ref{axial_I_Tc_Eta_root_n}\textcolor{blue}{(e)} (right \(y\)-axis). It is observed that \(\sigma_{F1}\) increases slightly from 1.72\% to 2.79\% as \(V_1\) increases from 0 to 5 V. This increase in \(\sigma_{F1}\) is a consequence of the enlarged effective size of the higher-order transparent zones with \(V_1\), as the size of the focused wave packet is limited by the width of the outermost zone \cite{PhysRevLett.83.4229,PhysRevA.91.043608}.

Therefore, biasing the FZP as \(V_n = V_1\sqrt{n}\) results in an increase in the transmission coefficient with the increase in \(V_1\). Furthermore, a maximum focusing efficiency of \(\eta_1 = 20.2\%\) is observed for the first time in He atom focusing using a biased FZP, which is twice the efficiency observed earlier using a Fresnel-Soret zone plate \cite{10.1119/1.18587, Greve:13}. Moreover, the first focal length (\(f_1\)) increases to \(1.463 f_{\text{cal}}\) while keeping the width of the focused wave packet \(\sigma_{F1}\) almost constant (within 3\%). These findings suggest improved controllability and enhanced efficiency in matter wave focusing by the FZP employing a biasing voltage that varies as \(V_n = V_1 \sqrt{n}\).
\subsubsection{$V_n=V_1 \sin(k_E n)$}
In this case, the radial modulation factor $k_E$ in the biasing voltage, along with its magnitude $V_1$, enhances the flexibility in the manipulation of wave packet by the FZP. The factor $k_E$ induces a periodic modulation in the polarization energy (Eq. \textcolor{blue}{(\ref{V_induced_along_zone_plate_root_N})}), which aids in controlling the transmission and phase of the wavefront emerging from the different zones.

The variation of the transmission coefficient $T_c$ with $k_E$ and $V_1$ is depicted as a contour plot in Fig. \ref{Tc_Eta_Axial_profile_Sin_kN}\textcolor{blue}{(a)}. It can be observed that, for lower values of $V_1$, the wave packet transmission remains almost independent of $k_E$, whereas for higher values of $V_1$ (e.g., $V_1=30$ V), $T_c$ exhibits a strong dependence on $k_E$. This is because, at smaller biasing voltages, the amplitude of the induced potential energy is so small that it does not have any significant effect on the wave packet transmission. A maximum transmission coefficient \( T_c = 0.323 \) is observed at \( V_1 = 10 \) V and \( k_E = 0.1 \), while a minimum value of \( T_c = 0.088 \) is observed at \( V_1 = 30 \) V and \( k_E = 0.2 \).

The variation of the first-order focusing efficiency \(\eta_1\) with \(k_E\) and \(V_1\) is shown in Fig. \ref{Tc_Eta_Axial_profile_Sin_kN}\textcolor{blue}{(b)}. It can be observed that the focusing efficiency strongly depends on both \(V_1\) and \(k_E\). A maximum value of \(\eta_1 = 10.49\%\) is observed at \(V_1 = 5\) V and \(k_E = 0.1\). Furthermore, to assess the intensity at F\textsubscript{1} for this case, the distribution of \(I_{X-\text{optical}}\) is shown in Fig. \ref{Tc_Eta_Axial_profile_Sin_kN}\textcolor{blue}{(c)} and compared with the \(I_{X-\text{optical}}\) obtained without biasing voltage (\(V_1 = 0\)). The peak intensity at F\textsubscript{1} for this case (\(V_1 = 5\) V and \(k_E = 0.1\)) is observed to be \(3.4 \times 10^{-2}\), which is the highest among all the cases with different \(k_E\) and \(V_1\).

The first focal length $f_1$ is found to vary from $0.89f_{cal}$ to $1.7f_{cal}$, and the size of the focused wave packet $\sigma_{F1}$ remains within 5\% of the width of the initial wave packet $\sigma_Y$. Therefore, the manipulation of wave packet transmission is controlled by the radial modulation factor \(k_E\) and biasing voltage \(V_1\) when the FZP is biased as \(V_n = V_1 \sin(k_E n)\). The induced periodic modulation in polarization energy allows control over wave packet transmission.

\section{Conclusion}\label{conclusions}

This study investigates the focusing of neutral helium atoms using a metallic Fresnel zone plate (FZP) by numerically solving the time-dependent Schrödinger equation with the split-step Fourier method. The FZP, made of chromium, is electrically biased in three ways: \(V_n = V_1\), \(V_n = V_1 \sqrt{n}\), and \(V_n = V_1 \sin (k_E n)\). Key findings include:

\begin{enumerate}
	\item Wave packet transmission increases with the temperature of incident particles, saturating at \(T_a \sim 0.642\) mK.
	\item Electrical biasing modulates the atom-FZP interaction, affecting diffractive focusing.
	\item Biasing at:
	\begin{enumerate}[label=\roman*.]
		\item \(V_n = V_1\) reduces transmission and focusing efficiency.
		\item \(V_n = V_1 \sqrt{n}\) enhances transmission by \(\sim 38\%\), increases focal length by $\sim73.5\%$, and improves focusing efficiency from 6\% to 20\%.
		\item \(V_n = V_1 \sin(k_E n)\) provides additional control through \(k_E\) and \(V_1\), achieving a focused wave packet with 10.5\% efficiency.
	\end{enumerate}
\end{enumerate}

The findings of this study will be helpful in manipulating the transmission and focal properties of an FZP employed in matter wave optics through external electrical biasing. The externally controlled focusing can enhance the efficiency of neutral helium-based microscopes employed in imaging sensitive surfaces. Furthermore, it introduces novel approaches for enhanced control in trapping cold atoms on atom chips and facilitates high-precision atom lithography for quantum electronic devices.

\nocite{*}
\bibliographystyle{apsrev4-2}
\bibliography{aipsamp}

\end{document}